\documentclass[aps,preprint,a4paper]{revtex4-1}
\usepackage{graphicx}
\usepackage{amsmath}
\usepackage{amssymb}
\usepackage{times}
\usepackage{ulem}
\usepackage[utf8]{inputenc}
\usepackage{placeins}
\usepackage[pdfborder=0]{hyperref}
\hypersetup{
    colorlinks=true,
    linkcolor=blue,          
    citecolor=blue           
}

\def\beq{\begin{equation}}
\def\eeq{\end{equation}}
\def\beqna{\begin{eqnarray}}
\def\eeqna{\end{eqnarray}}
\def\bea{\begin{array}}
\def\ea{\end{array}}

\newcommand{\etal}{\textit{et al.~}}
 
\begin{document}
\title{Frequency-comb response of a parametrically-driven Duffing oscillator to
a small added ac excitation}
\author{Adriano A. Batista$^1$}
\email{adriano@df.ufcg.edu.br}
\author{A.~A. Lisboa de Souza$^2$}
\affiliation{
$^1$Departamento de Física\\
Universidade Federal de Campina Grande\\
Campina Grande-PB,
CEP: 58109-970, Brazil\\
$^2$ Departamento de Engenharia Elétrica, Universidade Federal da Paraíba\\
João Pessoa-PB, CEP: 58.051-970, Brazil}
\date{\today}
\begin{abstract}
Here we present a one-degree-of-freedom model of a nonlinear
parametrically-driven resonator in the presence of a small added ac signal that
has spectral responses similar to a frequency comb.
The proposed nonlinear resonator has a spread spectrum response with a
series of narrow peaks that are equally spaced in frequency.
The system displays this behavior most strongly after a symmetry-breaking
bifurcation at the onset of parametric instability.
We further show that the added ac signal can suppress the transition to
parametric instability in the nonlinear oscillator.
We also show that the averaging method is able to capture the essential dynamics
involved. 
\end{abstract}
\maketitle
\section{Introduction}
The study of the effect of small signals on non-autonomous
nonlinear dynamical systems near bifurcation points was pioneered by 
K. Wiesenfeld and B. McNamara in the mid 80's \cite{wies85,wies86}.
They showed that several different dynamical systems are very sensitive to
coherent perturbations near the onset of codimention-one bifurcations,
such as period doubling, saddle node, transcritical, Hopf, and pitchfork
(symmetry-breaking) bifurcations.
They developed a general linear response theory, based on perturbation and
Floquet theories, explaining the effects of small coherent signals perturbing
limit cycles of nonlinear systems near the onset of bifurcation points.
One of the systems to which they applied their theory was the
ac-driven Duffing oscillator.
It was found by them that nonlinear dynamical systems could be used as
narrow-band phase sensitive amplifiers.

Parametric amplification has been studied in electronic systems since at least
from late 50's and early 60's by P. K. Tien \cite{tien1958parametric}, R.
Landauer \cite{landauer1960parametric}, and Louisell \cite{louis60}.
It has been used for its desirable characteristics of high gain and low noise
\cite{batista2011snr}.
Parametrically-driven Duffing oscillators have been used to model many different
physical systems such as the nonlinear dynamics of buckled beams 
\cite{nayfeh1989bifurcations, abou1993nonlinear}
or driven Rayleigh-Bénard convection \cite{lucke1985response}.
Homoclinic bifurcations were found in the
parametrically-driven Duffing oscillator \cite{parthasarathy1992homoclinic}.
Further bifurcations and chaos were found in \cite{jiang2017bifurcations}.
None of these papers investigated the type of spectral response we study here.

With the advent and development of micro-electromechanical systems (MEMS) technology in the 90's new
mechanical resonators were developed, such as the doubly clamped beam
resonators that could reach very high quality factors.
The dynamics of the fundamental mode of these resonators is well approximated by
the Duffing equation.
Furthermore, these micromechanical devices might exhibit, if properly tuned,  a
bistable response that can be quantitatively modeled by the bistability obtained
in Duffing oscillators \cite{aldridge05}.
More recently, amplifiers that operate 
near the threshold of bifurcations were shown to present very high-gain
amplification \cite{almog06, almog07, almog2007noise}.
A degenerate parametrically excited Duffing amplifier was proposed by
Rhoads \etal \cite{rhoads2010impact}.

One might be interested in a very sensitive high gain amplifier with a spread
spectrum, so as to sample selectively a broad band of frequencies.
This is particularly interesting for applications requiring spectrum sensing,
and one way to achieve this is through the use of frequency combs
\cite{jamali2019fully, yang2019broadband}.
The response of frequency combs to a narrowband small signal is given by
a series of narrow peaks equally spaced in frequency.
Cao \etal \cite{cao2014phononic} proposed a phononic frequency-comb generation
based on an ac driven Fermi-Pasta-Ulam-Tsingou chain.
In 2017, Ganesan et al \cite{ganesan2017phononic, ganesan2018phononic}
created the first mechanical resonators that had spectral lines
similar to optical frequency-combs.
Their experimental apparatus is based on two symmetrical cantilevers
mechanically coupled to one another at their bases.
They used two coupled nonlinear normal modes to model the dynamics in which the first mode
is resonantly excited by an added external drive, while the second mode is
parametrically excited by the first mode.
These experiments were followed by  Czaplewski et al
\cite{czaplewski2018bifurcation} in 2018,
who used another two-mode coupling nonlinear model to describe experimental data
from a mechanical resonator with both flexural and torsional vibrations.
In Ref.~\cite{houri2019modal}, the authors propose a theoretical model
based on an approximation to a nonlocal Euler-Bernoulli model with many
normal modes to describe frequency combs.
More recently, Singh \etal \cite{singh2020giant} observed a mechanical frequency
comb in a graphene-silicon nitride hybrid resonator.
They modeled the observed phenomena using two coupled normal modes, one
nonlinear (graphene) and the other linear (SiN).

We present one parametrically-driven Duffing oscillator model and make a 
nonlinear analysis of it based on the averaging method.
When there is no external added ac excitation, this nonlinear system can present
a bistable region (in which the quiescent solution and one limit cycle are
stable as seen in the original non-autonomous system).
From the perspective of the averaged equations one has a tristable region,
whose onset corresponds to a dual saddle-node bifurcation. 
There is also a threshold for parametric instability, which corresponds to a
pitchfork bifurcation in the averaged equations, either supercritical or subcritical.
With the application of an external ac drive, the nonlinear system may present
a frequency-comb-like behavior as the parametric pump amplitude is increased 
past the dual saddle-node bifurcation.
It occurs just after a sharp increase in the spectral
component corresponding to parametric instability.
This corresponds to a symmetry-breaking transition, in which the number of
spectral peaks is doubled with peaks equally spaced in the frequency spectrum.
All these responses are captured by the first-order averaging method.

In this article we show that a weaker form of mechanical frequency combs (MFCs)
gradually appears when the parametric pump amplitude is increased.
In addition to that, a stronger form of MFCs occurs after a symmetry-breaking
bifurcation.
In this form, there are twice as many peaks in the Duffing oscillator response
to the added ac excitation as in the weaker form of the frequency comb.
Furthermore, unlike the models proposed in Refs. \cite{ganesan2017phononic,
ganesan2018phononic, czaplewski2018bifurcation}, we show that only one
parametrically-driven nonlinear mode is needed to present the
frequency-comb-like behavior in the spectral response of the resonator.
Different from the analysis developed by Bryant and Wiesenfeld
\cite{bryant1986suppression}, there is no typical period-doubling bifurcation 
here. 
\section{The parametrically-driven Duffing oscillator model}
\label{nonLinearOPF}
\begin{figure}[!ht]
    \centering{\includegraphics[scale=0.7]{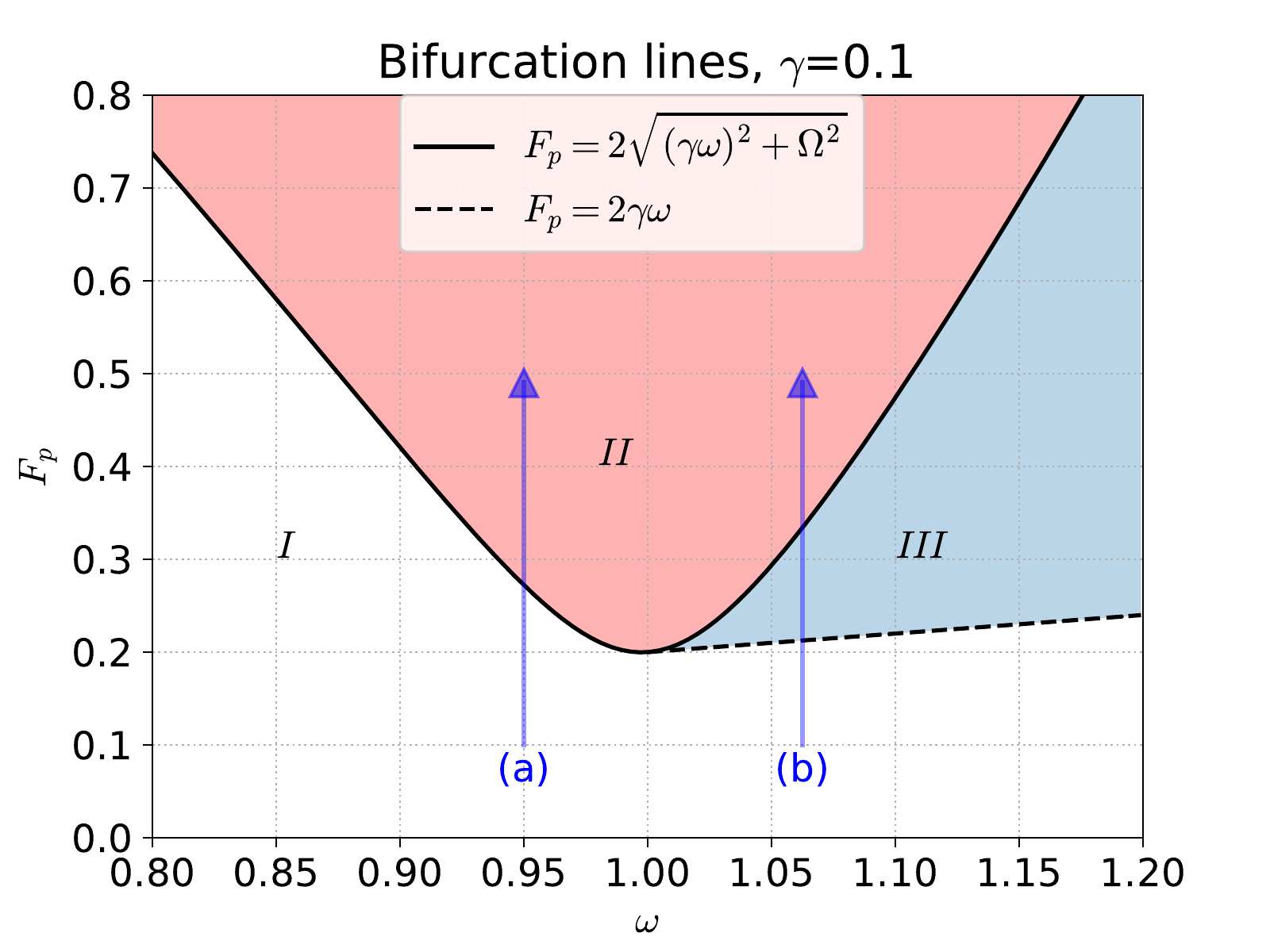}}
    \caption{Bifurcation lines of the averaged equations \eqref{POavg1} of the
    parametrically-driven Duffing oscillator. 
    In region I there is only one fixed point--the quiescent solution--which is
    stable.
    In region II, above the black continuous line, there are three fixed points:
    the unstable quiescent solution (a saddle point) and two stable nodes or
    spirals. When the continuous line is crossed, as $F_p$ is increased, a
    supercritical pitchfork bifurcation occurs for $\omega<1$, whereas for
    $\omega>1$, a subcritical pitchfork bifurcation occurs. 
    On the dashed line a dual saddle-node bifurcation occurs.
    In region III we have five fixed points: the stable quiescent solution (a
    spiral or a node), two saddle points, and two stable nodes, or stable
    spirals.} 
    \label{fig:bifurcationLines} 
\end{figure}

The one-degree of freedom model we use to describe the dynamics of a
parametrically-driven nonlinear resonator is given by
\beq
\ddot x(t)=-\gamma \dot x(t)-x(t)-\alpha x^3(t)+F_p\cos(2\omega t)x(t)
\label{oscParametricoNLinear}
\eeq
in dimensionless units.
Here $\gamma$ is the dissipation rate, $\alpha$ is the nonlinear coefficient,
$F_p$ is the parametric pump amplitude, and $2\omega$ is the parametric pump
angular frequency.
Assuming $\gamma, \alpha,$ and $F_p$ are $O(\epsilon)$, with $0<\epsilon<<1$,
we can apply the averaging method \cite{Guck83} to obtain a slow autonomous dynamics.
This is accomplished via the transform 
\beq
\begin{aligned}
x(t) &=u(t)\cos(\omega t)-v(t)\sin(\omega t),\\
\dot x(t) &=-\omega\left[u(t)\sin(\omega t)+v(t)\cos(\omega t)\right].
\end{aligned}
\label{avgTrans}
\eeq
After applying this change of variables and neglecting fast oscillating terms, 
via Poincaré weakly non-linear transformation, we obtain
\beq
\begin{aligned}
\dot {u}&= -\frac{1}{2 \omega}\bigg\{\gamma\omega u
+\left[\Omega+\frac{F_p}{2}+\frac{3\alpha}{4}(u^{2}+v^2)\right]v\bigg\}, \\
\dot {v}&=-\frac{1}{2
\omega}\left\{\left[-\Omega+\frac{F_p}{2}-\frac{3\alpha}{4}(u^{2}+v^2)\right]u+\gamma\omega
v \right\},
\end{aligned}
\label{POavg1}
\eeq
where $\Omega=1-\omega^2=O(\epsilon)$.
The fixed points are obtained from the solution of
\beq
\begin{aligned}
    &\gamma\omega u +\left[\Omega+\frac{F_p}{2}+\frac{3\alpha}{4}r^2\right]v=0, 
\\
    &\left[-\Omega+\frac{F_p}{2}-\frac{3\alpha}{4}r^2\right]u+\gamma\omega v =0,
\end{aligned}
\label{eq:fixedPt1}
\eeq
where $r^2=u^2+v^2$.
The characteristic equation based on Eq.~\eqref{eq:fixedPt1} can be written as
\beq
\frac{F_p^2}{4}=(\gamma\omega)^2+\left(\Omega+\frac{3\alpha}{4}r^2\right)^2\\
\eeq
The steady-state squared amplitude is given by
\beq
r^2=-\frac{4}{3\alpha}\left[\Omega\pm\sqrt{\frac{F_p^2}{4}-\gamma^2\omega^2}\,\right].
\label{r2}
\eeq
In Fig.~\ref{fig:bifurcationLines}, we present the bifurcation diagram in the
$\omega\times F_p$ parameter space of the averaged equations of motion,
Eq.~\eqref{POavg1}, of the parametrically-driven Duffing
oscillator.
When $\omega<1$ ($\Omega>0$) and $|F_p|<2\sqrt{(\gamma\omega)^2+\Omega^2}$
the only possible solution is $r=0$.
When $\omega>1$ ($\Omega<0$) and $|F_p|<2\gamma\omega$, we also find the fixed-point solution with $r=0$.
These conditions characterize region I of the bifurcation diagram.
In region II, we have $|F_p|>2\sqrt{(\gamma\omega)^2+\Omega^2}$.
In this region, the steady-state amplitude $r$ admits two solutions: $r=0$ and
\beq
r_+=\sqrt{\frac{4}{3\alpha}\left(-\Omega+\sqrt{\frac{F_p^2}{4}-\gamma^2\omega^2}\,\right)}.
\label{r+}
\eeq
With $u=r\cos\theta$ and $v=r\sin\theta$, we obtain the steady-state angle
$\theta$ from
\[
\tan\theta_+=-\dfrac{\gamma\omega}{\Omega+F_p/2+\frac{3\alpha r^2_+}4}.
\]
In region III, we have $\Omega<0$ and $2\gamma\omega<|F_p|<2\sqrt{(\gamma\omega)^2+\Omega^2}$.
In this region, there are three different solutions for the steady-state 
amplitude $r$:
$0$, and
\beq
r_\pm=\sqrt{-\frac{4}{3\alpha}\left(\Omega\pm\sqrt{\frac{F_p^2}{4}-\gamma^2\omega^2}\,\right)}.
\label{r_pm}
\eeq
The corresponding values of the steady-state angle $\theta$ can be obtained from
\[
\tan\theta_\pm=-\dfrac{\gamma\omega}{\Omega+F_p/2+\frac{3\alpha r^2_\pm}4}.
\]
We determine the stability of the fixed points based on the eigenvalues of
the Jacobian matrix.
At a fixed point $(\bar{u}, \bar{v})$, it is given by
\beq
\begin{aligned}
Df &=\left(\begin{array}{cc}
\frac{\partial f_{1}}{\partial u}&\frac{\partial f_{1}}{\partial v}\\
\frac{\partial f_{2}}{\partial u}&\frac{\partial f_{2}}{\partial v}\\
\end{array}
\right)_{(\bar{u},\bar{v})}\\
&=-\frac{1}{2\omega}\left(\begin{array}{cc}
\gamma\omega+\frac{3\alpha}{2}\bar u\bar
v&\Omega+\frac{F_p}{2}+\frac{3\alpha}{4}(\bar u^2+3\bar v^2)\\
-\Omega+\frac{F_p}{2}-\frac{3\alpha}{4}(3\bar u^2+\bar v^2)
&\gamma\omega-\frac{3\alpha}{2}\bar u\bar v\\
\end{array}\right). 
\end{aligned}
\eeq
For the $(0,0)$ fixed point, we obtain
\beq
Df(0,0)=-\frac{1}{2\omega}\left(\begin{array}{cc}
\gamma\omega
&\Omega+\frac{F_p}{2}\\
-\Omega+\frac{F_p}{2}&\gamma\omega\\
\end{array}\right). 
\eeq
The eigenvalues are given by
\beq
\lambda_\pm=-\frac\gamma2\pm\frac1{2\omega}\sqrt{\frac{F_p^2}4-\Omega^2}\,
\eeq
Note that when $|F_p|<2\Omega$, the eigenvalues are complex and the quiescent
mode fixed-point $(0,0)$ is a stable spiral.
When $2\Omega<|F_p|<2\sqrt{(\gamma\omega)^2+\Omega^2}$, the two eigenvalues 
are real and this fixed point is a stable node.
Hence, in regions I and III, this fixed point is stable, whereas, in region II,
it is unstable.
At the transition line to parametric instability, the stable node becomes an
saddle point as one goes from either region I or III into region II.

\begin{figure}[!ht]
    \centering{\includegraphics[scale=0.5]{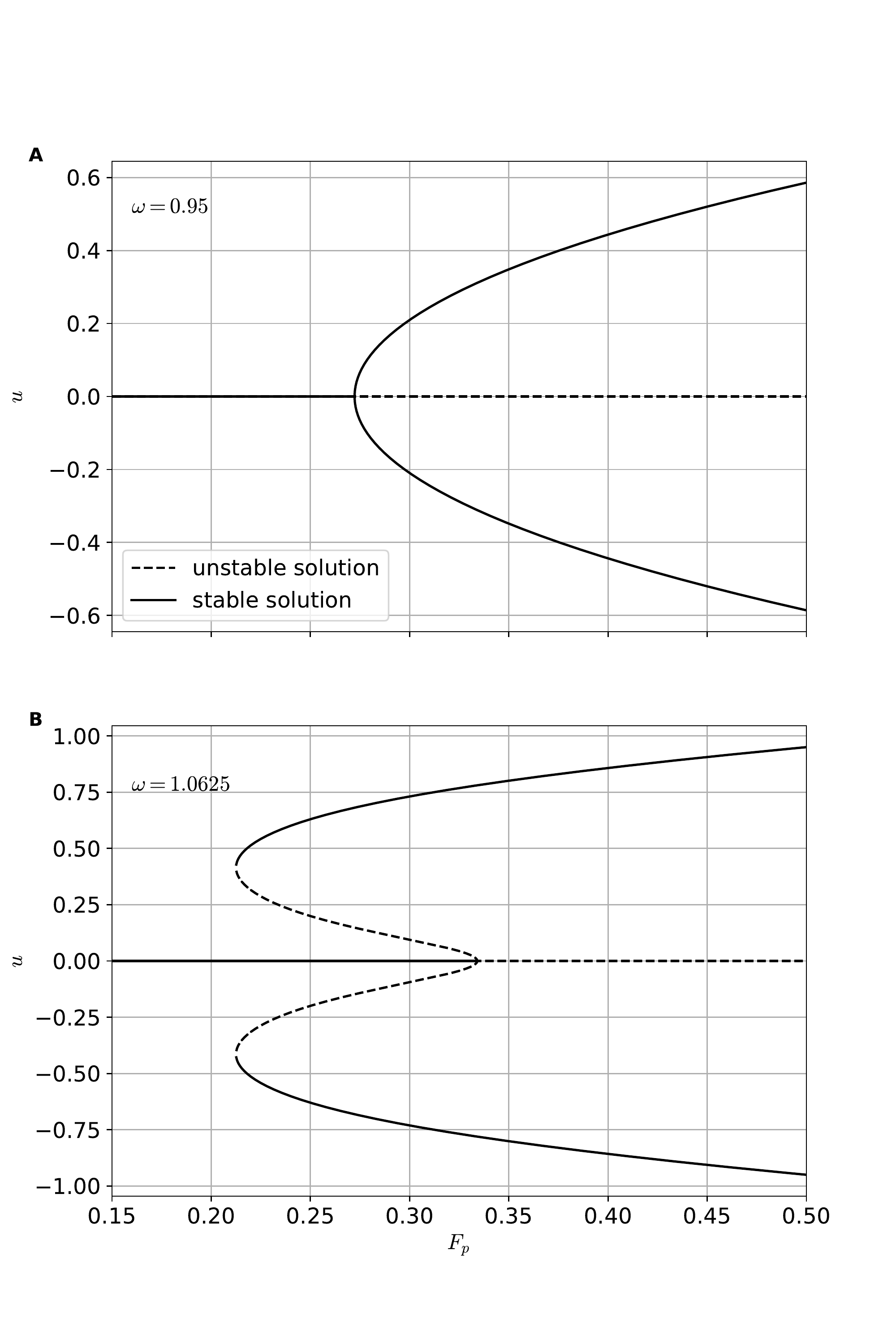}}
    \caption{
    In Frame {\bf A} we can see a supercritical pitchfork bifurcation at
    approximately $F_p=0.272259$.
    In Frame {\bf B} we can see a double saddle-node bifurcation at $F_p=0.2125$ and
    a subcritical pitchfork bifurcation at approximately $F_p=0.334101$.
    }
    \label{fig:uResp} 
\end{figure}

In Fig.~\ref{fig:uResp} we plot the bifurcation diagrams along lines (a) and (b)
of Fig.~\ref{fig:bifurcationLines}.
In frame A, we obtain a supercritical pitchfork bifurcation as one crosses the
parametric instability threshold along line (a).
In frame B, we obtain a dual saddle-node bifurcation at about $F_p=0.2125$,
and a subcritical pitchfork bifurcation at $F_p=0.334101$ along line (b).

\FloatBarrier

\section{The Duffing parametric amplifier}
If we add to Eq.~\eqref{oscParametricoNLinear} an ac excitation, 
we obtain a parametric amplifier. 
Here we call it a Duffing amplifier (DA).
It is described by the equation
\begin{eqnarray}
\ddot{x}(t)&=&-\gamma \dot{x}(t)-x(t)-\alpha x^{3}(t)+F_{p}\cos(2\omega t)x(t)+F_s\cos(\omega_s t+\varphi_{0}),
\label{eq:ampParamNaoLin1}
\end{eqnarray}
where $F_s$ is the amplitude, $\omega_s$ is the angular frequency,
and $\varphi_0$ is an arbitrary phase of the external ac excitation.
Assuming that $F_s=O(\epsilon)$, $\delta=\omega_s-\omega=O(\epsilon)$ and all the
other coefficients are also small as in the previous section, we can apply
the averaging method.
After doing so, we find the slowly-varying dynamics
\beq
\begin{aligned}
\dot {u} &=-\frac{1}{2 \omega}\left\{\gamma\omega u
+\left[\Omega+\frac{F_p}{2}+\frac{3\alpha}{4}(u^{2}+v^2)\right]v\right\}+\frac{F_s}{2\omega}\sin\varphi(t),\\
\dot {v} &=-\frac{1}{2
\omega}\bigg\{\left[-\Omega+\frac{F_p}{2}-\frac{3\alpha}{4}(u^{2}+v^2)\right]u+\gamma\omega v
\bigg\}-\frac{F_s}{2\omega}\cos\varphi(t),
\end{aligned}
\label{mediaVAm1}
\eeq
where $\varphi(t)=\delta t+\varphi_0$.

\section{Results and discussion}
\begin{figure}[!ht]
\centering{\includegraphics[scale=0.8]{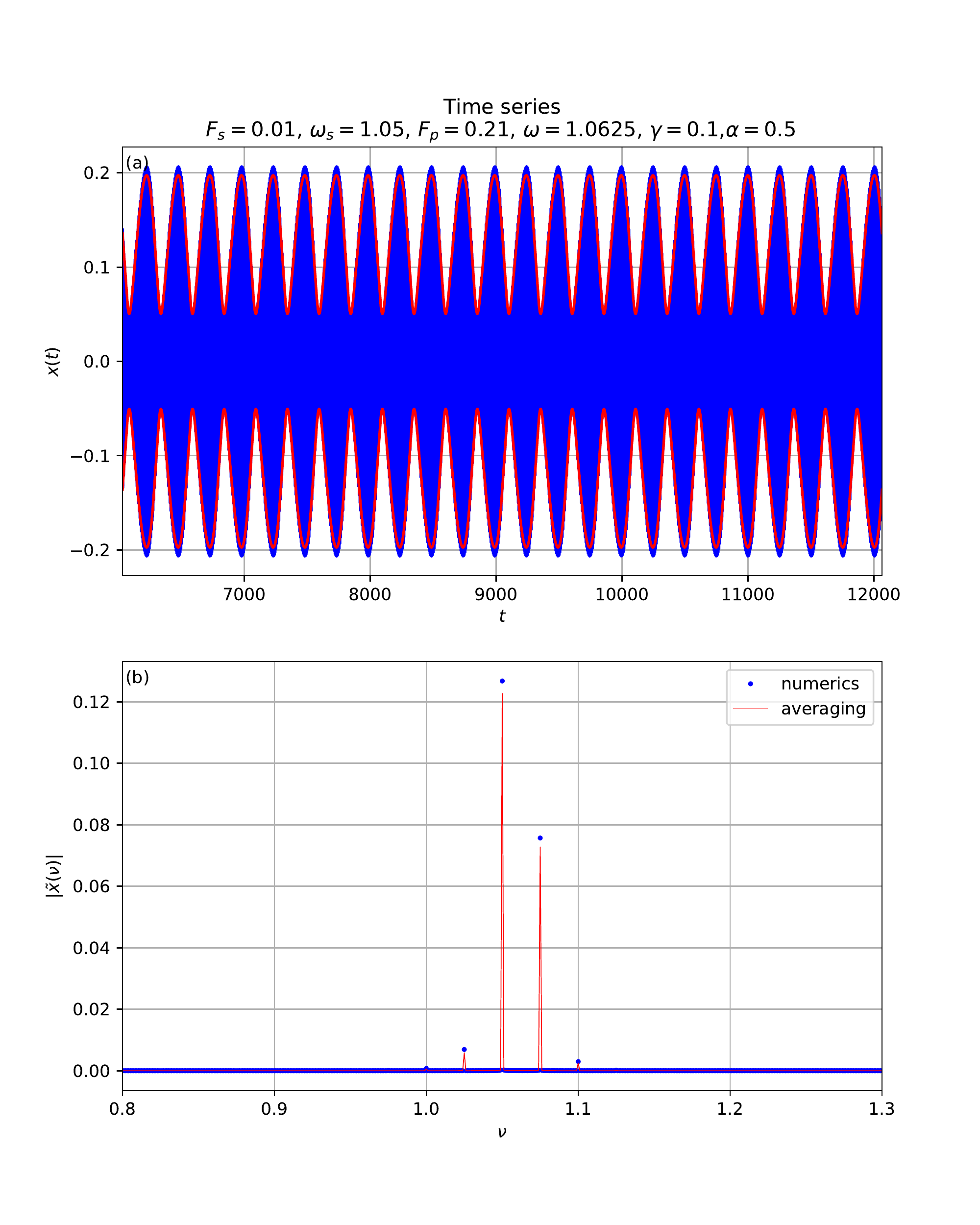}}
\caption{
(a) Time series of the Duffing amplifier (DA) obtained from the numerical integration of
Eq.~\eqref{eq:ampParamNaoLin1}.\\
(b) The corresponding Fourier Transform. 
The strongest peak is the signal, at $\nu=1.05$, whereas the second strongest
peak is the idler.
These two peaks (signal and idler) are hallmarks of a parametric amplifier.
}
\label{fig:tSeriesFTFreqComb0}
\end{figure}

\begin{figure}[!ht]
\centering{\includegraphics[scale=0.8]{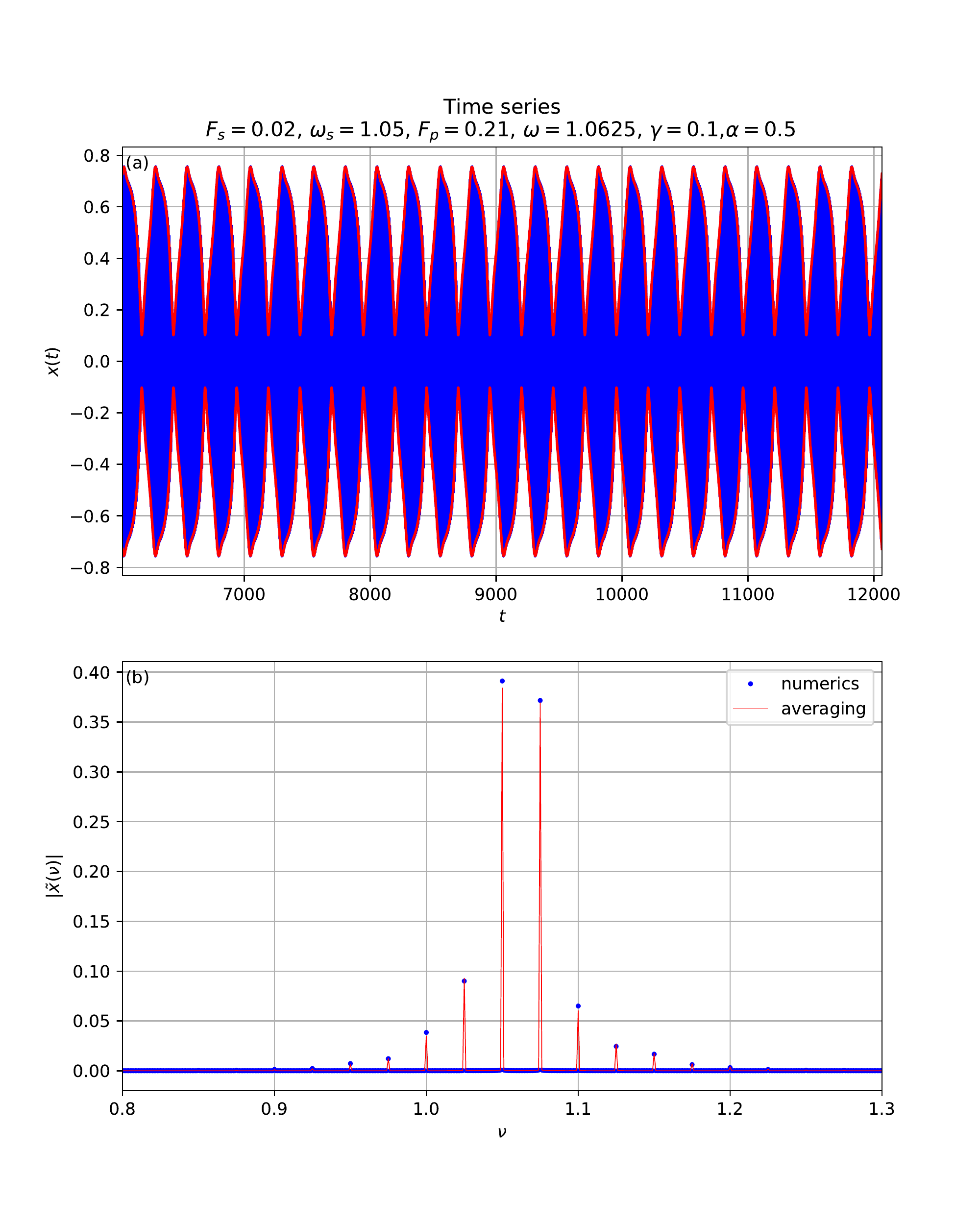}}
\caption{
(a) Time series of the Duffing amplifier (DA) obtained from the numerical integration of
Eq.~\eqref{eq:ampParamNaoLin1}.\\
(b) The corresponding Fourier Transform. 
One can see a frequency-comb-like behavior.
The strongest peak is the signal, at $\nu=1.05$, whereas the second strongest
peak is the idler.
The other peaks are due to the nonlinear nature of the amplifier.
There are peaks only at $\omega\pm(2n+1)\delta$, where $n=0,1,2,\dots$.
}
\label{fig:tSeriesFTFreqComb1}
\end{figure}

\begin{figure}[!ht]
\centering{\includegraphics[scale=0.8]{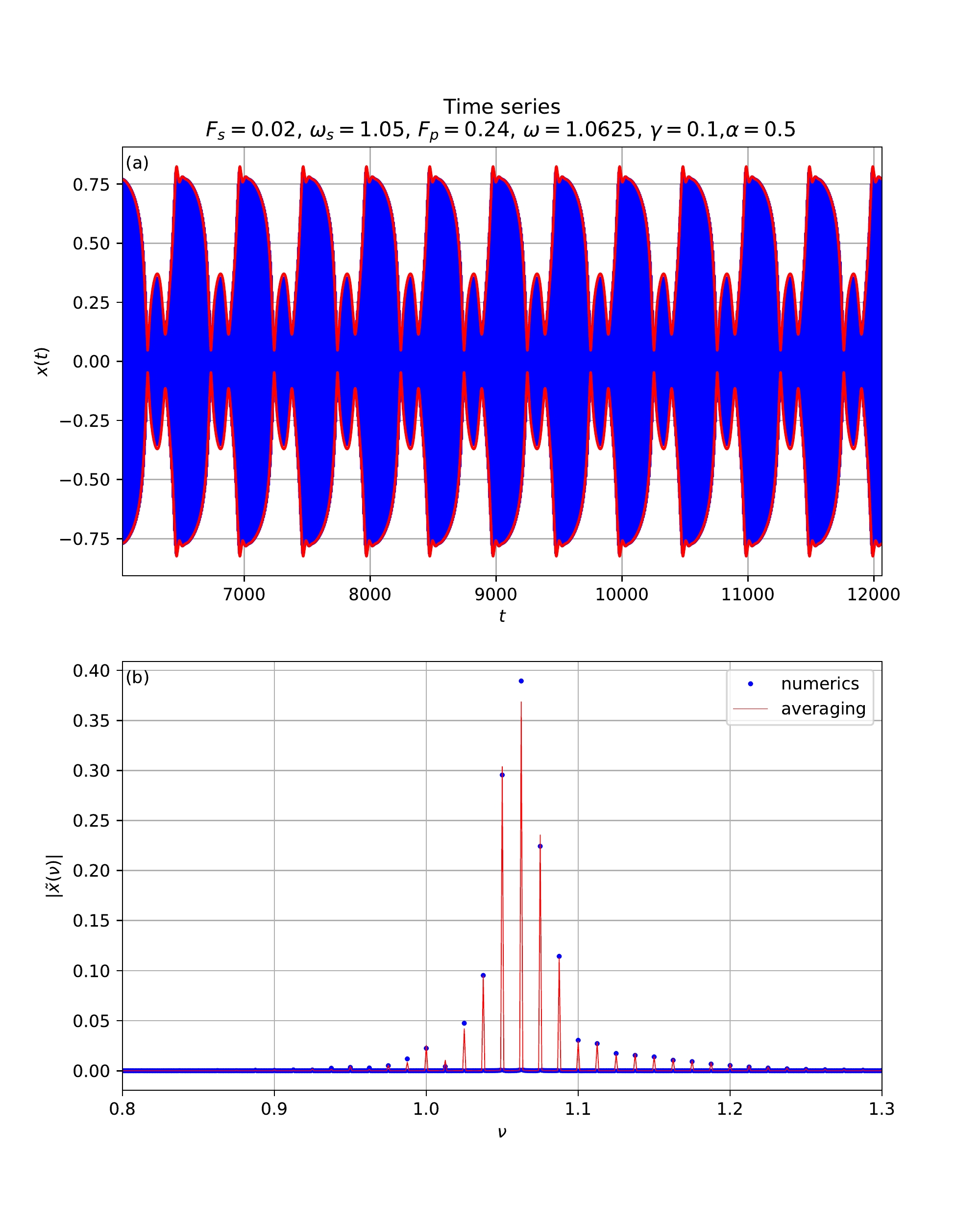}}
\caption{
(a) Time series of the DA obtained from the numerical integration of
Eq.~\eqref{eq:ampParamNaoLin1}. 
(b) The corresponding Fourier Transform.
One can see a stronger frequency-comb behavior with roughly twice as many
peaks as in Fig.~\ref{fig:tSeriesFTFreqComb1}(b).
}
\label{fig:tSeriesFTFreqComb1n}
\end{figure}

\begin{figure}[!ht]
    \centering{\includegraphics[scale=0.8]{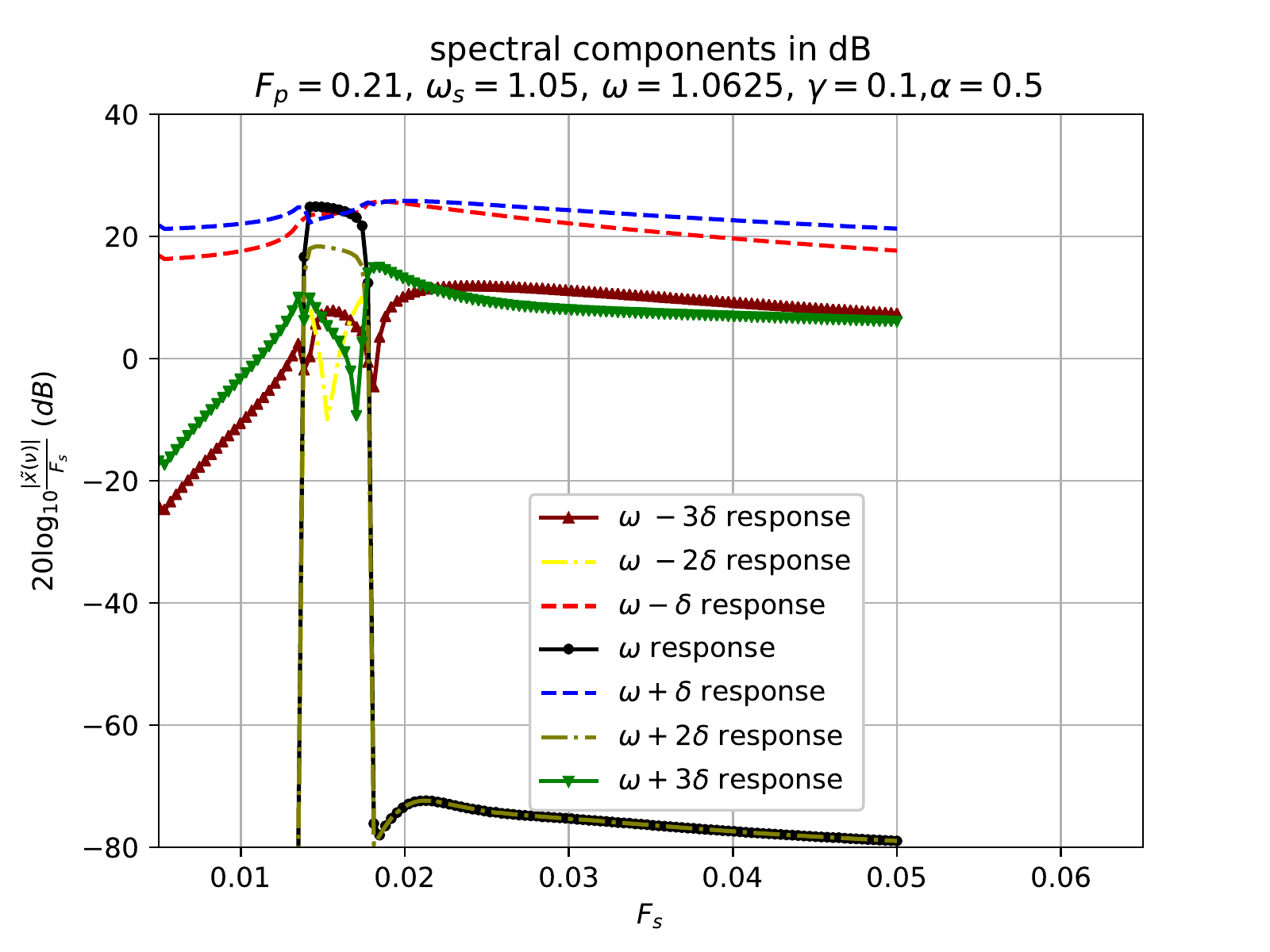}}
    \caption{ 
    A parametric instability occurs in the interval $0.01386<F_s<0.01811$.
    This response appears due to a symmetry-breaking bifurcation at the limits
    of this interval.
    }
    \label{fig:symBreak}
\end{figure}

\begin{figure}[!ht]
    \centering{\includegraphics[scale=0.8]{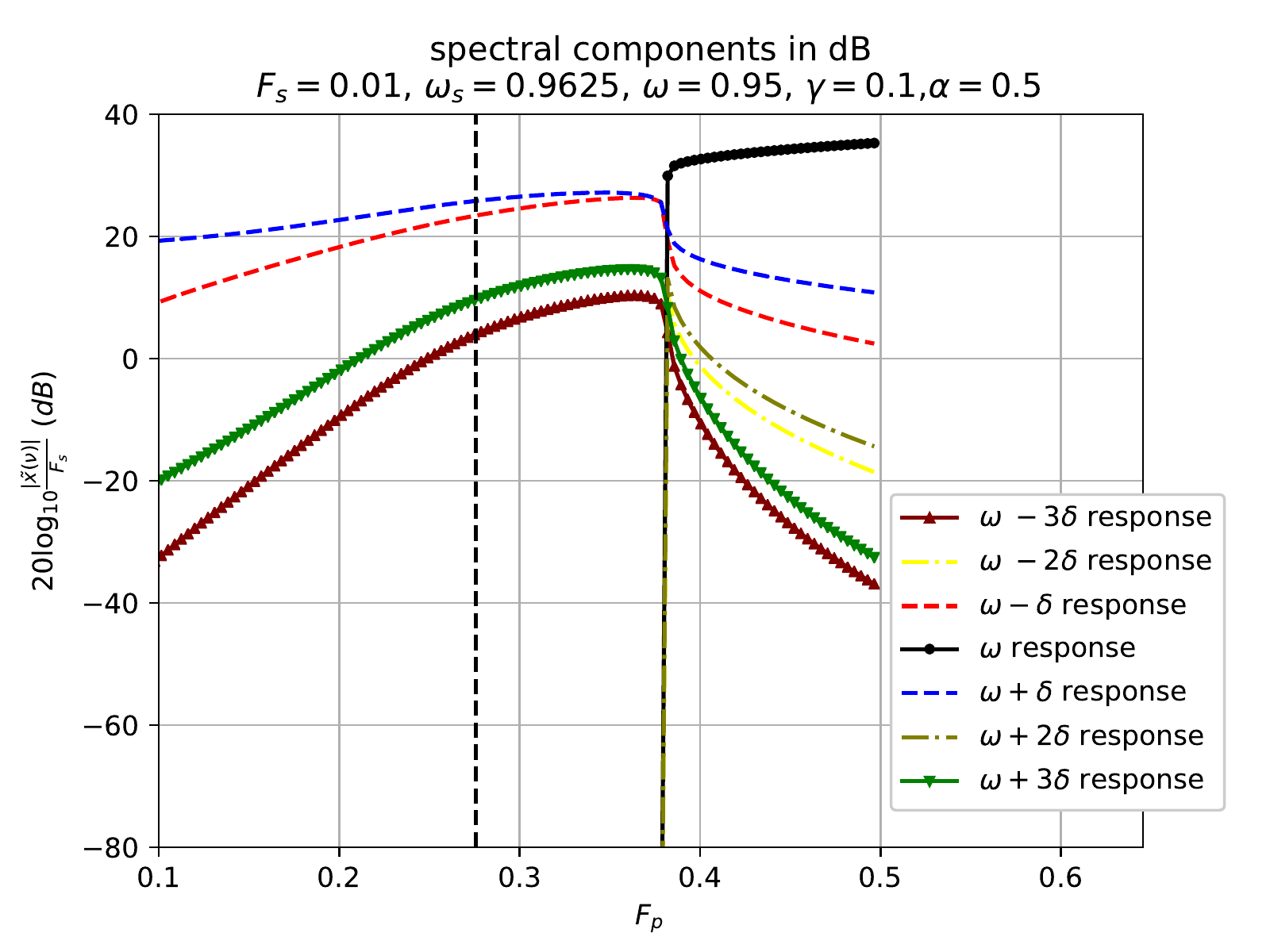}}
    \caption{Suppression of parametric instability.
    Here we plot the seven main spectral peaks of the FT as a function of pump
    amplitude $F_p$.
    In parameter space, the variation of $F_p$ occurs along line $(a)$ of
    Fig.~\ref{fig:bifurcationLines}, although the system investigated here
    obeys Eq.~\eqref{eq:ampParamNaoLin1}, in which there is an added ac excitation.
    The vertical dashed line indicates the transition to parametric instability.
    The suppression is measured by the amplitude of the spectral peak of the
    Fourier transform $|\tilde
    x(\nu)|$ at $\nu=\omega$.
    Here the frequency-comb-like behavior reaches a maximum when the peak at
    $\omega$ is minimal approximately when the parametric pump amplitude $F_p=0.36$.
    }

\label{fig:paramInstSup1}
\end{figure}

\begin{figure}[!ht]
    \centering{\includegraphics[scale=0.8]{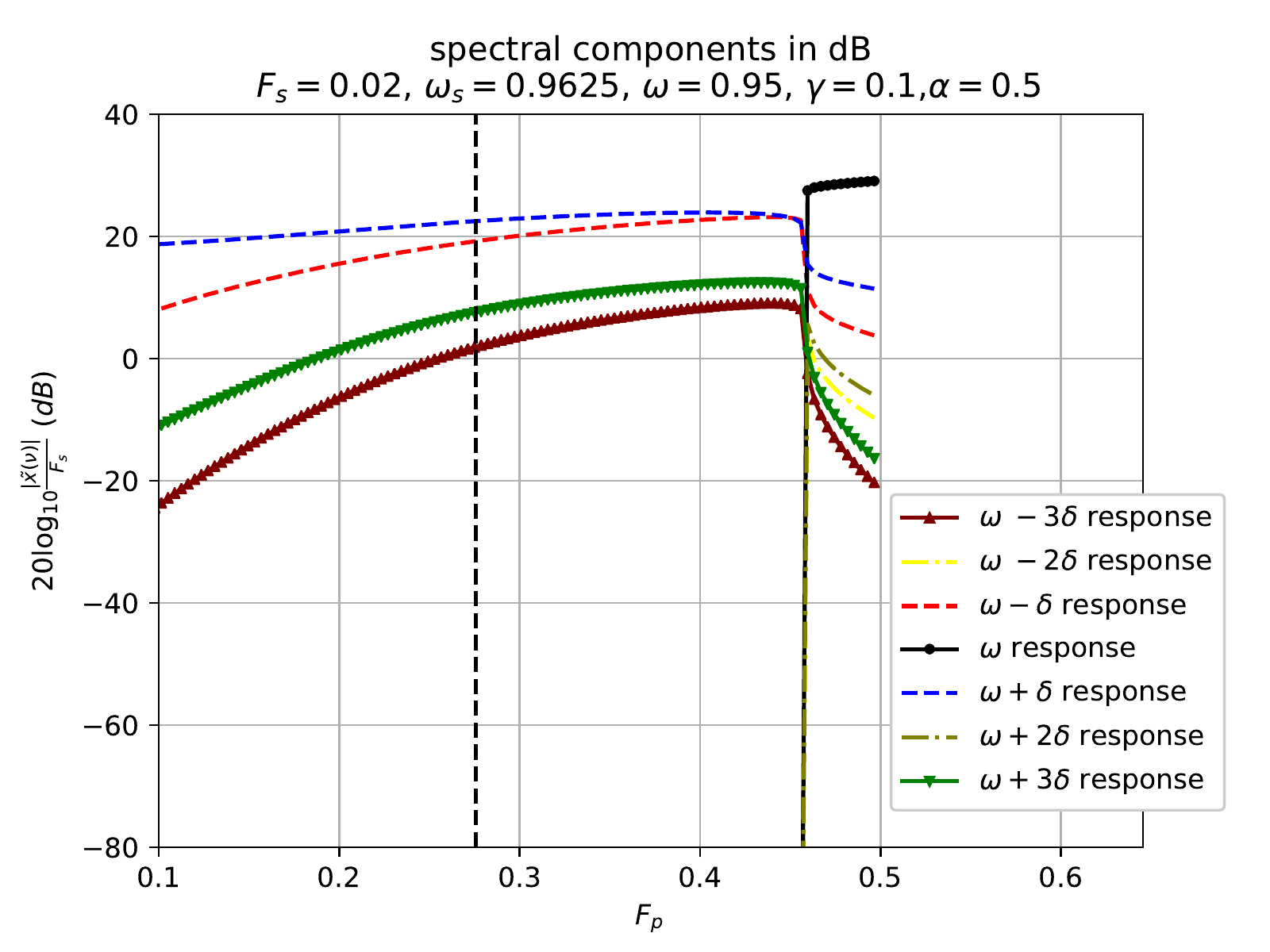}}
    \caption{Suppression of parametric instability.
    The vertical dashed line indicates the transition to parametric instability.
    Note the increased suppression of parametric instability for larger value of
    external signal amplitude as compared with the
    results of previous figure.
    }
\label{fig:paramInstSup1n}
\end{figure}

\begin{figure}[!ht]
    \centering{\includegraphics[scale=0.8]{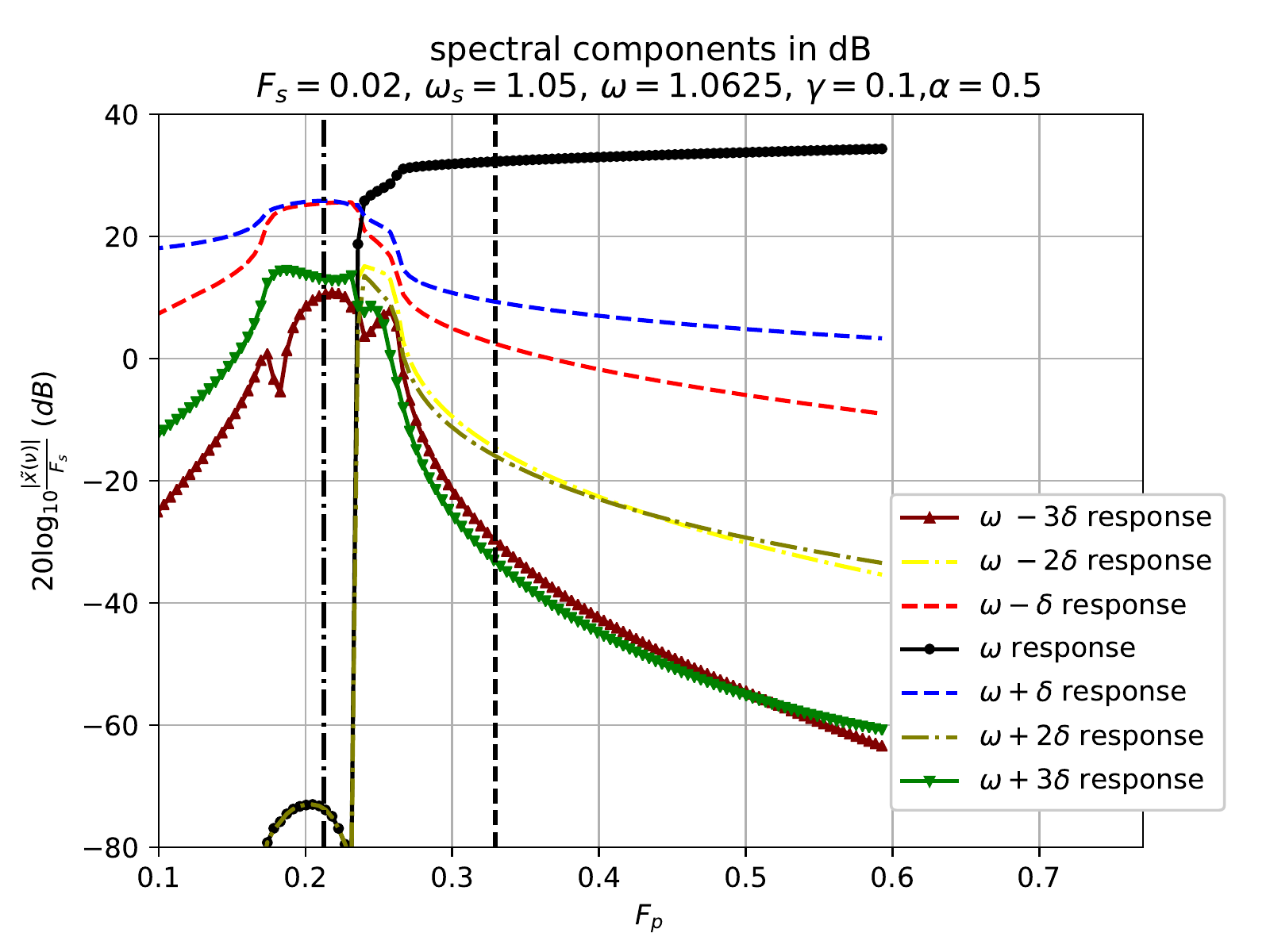}}
    \caption{Broader frequency comb generation. The strongest frequency-comb behavior roughly occurs in the
    range $0.23<F_p<0.25$.
    The vertical dashed-dotted line indicates the dual saddle-node bifurcation 
    and the vertical dashed line indicates the parametric instability
    transition in the parametrically-driven Duffing oscillator.
    }
\label{fig:freqCombPeaks1}
\end{figure}

\begin{figure}[!ht]
    \centering{\includegraphics[scale=0.8]{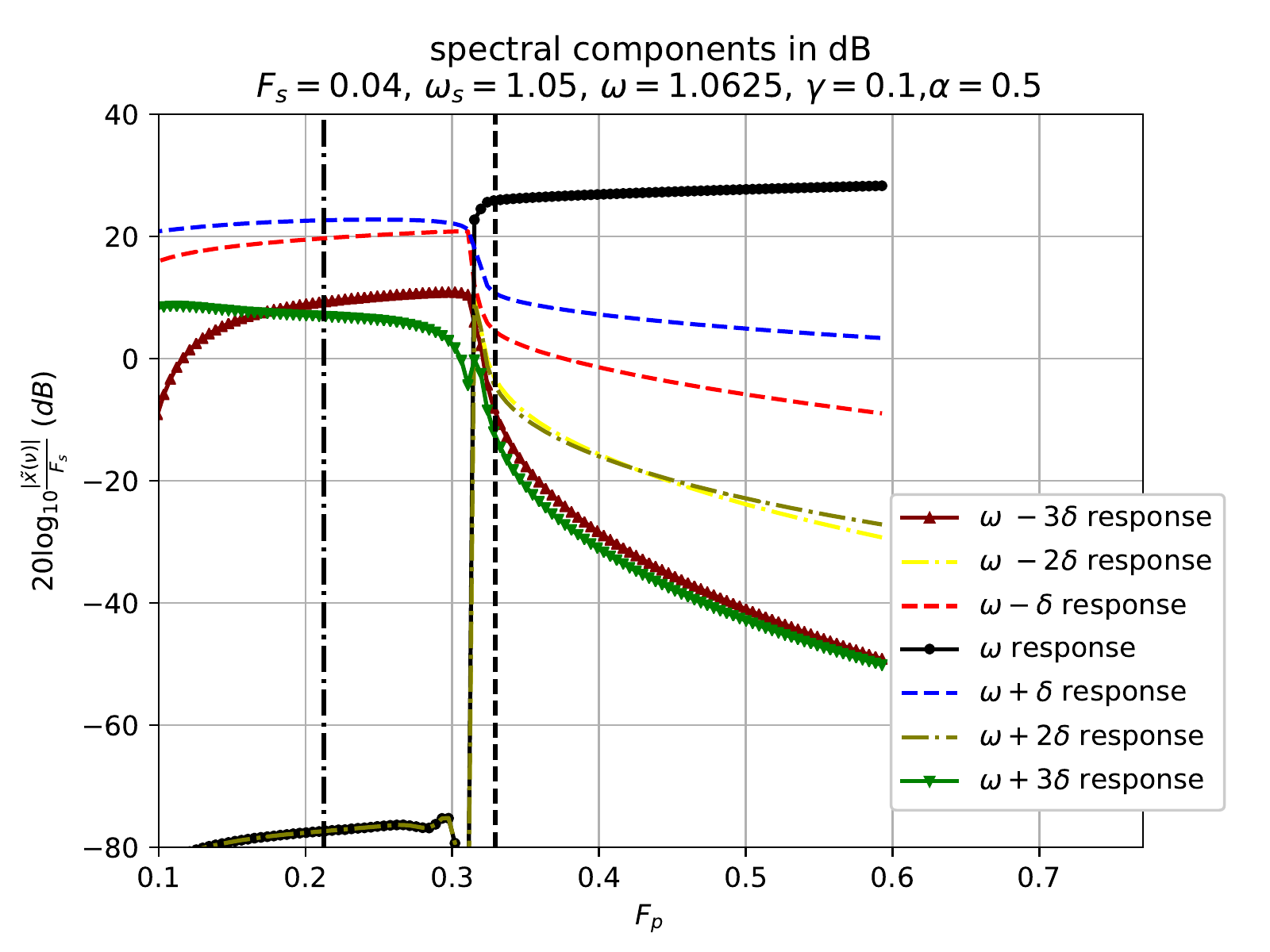}}
    \caption{Very narrow frequency comb generation. Note the suppression of the
    parametric instability transition as the amplitude of the ac excitation
    $F_s$ increases. The transition still occurs in region III of the
    bifurcation diagram of Fig. 1, but very
    near the parametric instability threshold (dashed line).
    }
\label{fig:freqCombPeaks2}
\end{figure}

We used the Odeint function of the Python’s scientific library package SciPy
\cite{scipy1.0} to integrate Eqs.~\eqref{eq:ampParamNaoLin1} and \eqref{mediaVAm1}.
The integration time-step used in each numerical result was
$h=\frac T{512}$, where $T=\frac{2\pi}{\omega}$.
To avoid end discontinuities in the the time series we chose the total time
of integration to be an integer multiple of $T_p=\frac{\omega
T}{|\delta|}=\frac{2\pi}{|\delta|}$.
For each time series in which we performed the Fourier transform (FT) (Figs.
\ref{fig:tSeriesFTFreqComb0}-\ref{fig:tSeriesFTFreqComb1n}), we run the equations of
motion for a time interval of $24T_p$, of which we discarded
the first half as an equilibration time so that all transients die out.
The fast Fourier transform routine used was SciPy's fftpack.
In Figs \ref{fig:symBreak}-\ref{fig:freqCombPeaks2},
we calculated the peak amplitudes using the following method
\cite{rajamani2017variation}  
\beq
\begin{aligned}
    a_n &= \frac2{\Delta t}\int_{t_0}^{t_0+\Delta t} x(t)\cos((\omega+n\delta)t)dt,\\
    b_n &= \frac2{\Delta t}\int_{t_0}^{t_0+\Delta t} x(t)\sin((\omega+n\delta)t)dt,\\
\end{aligned}
\eeq
hence the amplitude of each peak is $r_n=\sqrt{a_n^2+b_n^2}$.
This works because we have approximately
$x(t)\approx\sum_{n=-N}^N\left[a_n\cos((\omega +n\delta)t)+b_n\sin((\omega
+n\delta)t)\right]=\sum_{n=-N}^Nr_n\cos((\omega +n\delta)t-\varphi_n)$. 
Here $2N+1$ is the total number of visible peaks of the comb after the
symmetry-breaking bifurcation,
whereas before the symmetry-breaking bifurcation only peaks at odd values of $n$
contribute.

 \FloatBarrier
In frame (a) of Fig.~\ref{fig:tSeriesFTFreqComb0}, we show a time series of numerical integration
of Eq.~\eqref{eq:ampParamNaoLin1}, which corresponds to parametric amplification
in the DA with the parametric pump set at $F_p=0.21$ and the ac excitation
amplitude set at $F_s=0.01$.
The  envelopes are obtained from the averaged slowly-varying dynamics given in
Eqs.~\eqref{mediaVAm1}.  
The envelope of the pulses is approximately symmetric in time.
In frame (b) of Fig.~\ref{fig:tSeriesFTFreqComb0}, we plot the FT corresponding to the time series shown
in frame (a). 
The semi-analytic approximation to the numerical FT spectrum was obtained from
the numerical integration of the corresponding averaged system, given in
Eqs.~(\ref{mediaVAm1}), with the transformation defined in Eq.~\eqref{avgTrans}.
The first-order harmonic balance approximation is roughly accurate, since it only
predicts the signal ($\omega_s=\omega+\delta$) and idler ($\omega-\delta$) peaks in a parametric amplifier.
Here, besides the signal  and idler 
peaks, we barely see two other sidebands due to the nonlinearity of our system.

In frame (a) of Fig.~\ref{fig:tSeriesFTFreqComb1}, we show a time series of numerical integration
of Eq.~\eqref{eq:ampParamNaoLin1}, which corresponds to parametric amplification
in the DA with the parametric pump set at $F_p=0.21$ and the ac excitation
amplitude set at $F_s=0.02$.
The  envelopes are obtained from the averaged slowly-varying dynamics given in
Eqs.~\eqref{mediaVAm1}.  
Note the increased time-inversion asymmetry in the envelope.
In frame (b) of Fig.~\ref{fig:tSeriesFTFreqComb1}, we plot the Fourier transform (FT) corresponding to the time series shown
in frame (a). 
The semi-analytic approximation to the numerical FT spectrum was obtained from
the numerical integration of the corresponding averaged system, given in
Eqs.~(\ref{mediaVAm1}), with the transformation defined in Eq.~\eqref{avgTrans}.
The first-order harmonic balance approximation becomes inaccurate, since it only
predicts the signal and idler peaks in a parametric amplifier.
Here, besides the signal and idler 
peaks, we also have sidebands spaced from one another by $2\delta$ 
and symmetrically positioned around these two central peaks.
One can clearly see a frequency-comb-like spectrum with ten easily seen peaks.

In frame (a) of Fig.~\ref{fig:tSeriesFTFreqComb1n}, with a higher value of the parametric pump
amplitude ($F_p=0.24$), the envelope of the DA response becomes even more asymmetric.
One can see that for the same time span the number of pulses is halved as
compared to the time series of Fig.~\ref{fig:tSeriesFTFreqComb1}.
This could indicate that there was a period-doubling bifurcation in the dynamical
system of Eq.~\eqref{mediaVAm1}, when the pump amplitude was increased from
$F_p=0.21$, in Fig.~\ref{fig:tSeriesFTFreqComb1}(a), to $F_p=0.24$.
In reality there is no period-doubling bifurcation there.
In frame (b) of Fig.~\ref{fig:tSeriesFTFreqComb1n}, in addition to the signal and idler peaks, we also have a strong peak at $\omega$, 
which is at half the parametric drive frequency.
Further new peaks can be seen at $\omega\pm2\delta$, $\omega\pm4\delta$,
$\omega\pm6\delta$, \dots.
If one increases further the parametric drive amplitude, then one gets a full
transition to parametric instability in which one gets a very strong peak at
$\nu=\omega$ with two small sidebands, the signal and the idler. 
The other spectral peaks of the frequency-comb tend to decrease gradually.

In Fig.~\ref{fig:symBreak}, we show that a parametric instability occurs in a narrow
window of amplitudes of the added ac excitation.
In this region occurs the strongest forms of frequency-comb response such as
occurs in Fig.~\ref{fig:tSeriesFTFreqComb1n}.
For $F_s< 0.01386$ the solutions are more symmetrical such as in
Fig.~\ref{fig:tSeriesFTFreqComb0}, inside the window they
are least symmetrical.
Whereas for $F_s>0.0181$, there is parametric instability suppression, less
asymmetry, and the frequency-comb response is weaker, with peaks spaced by
$2\delta$ (see Fig.~\ref{fig:tSeriesFTFreqComb1}).

In Figs. \ref{fig:paramInstSup1}-\ref{fig:freqCombPeaks2}, we show how the seven main
spectral peaks vary as a function of the parametric pump amplitude.
In Figs. \ref{fig:paramInstSup1}-\ref{fig:paramInstSup1n}, we fix $\omega$ at $0.95$ and
we increase $F_p$ along line $(a)$ of the bifurcation diagram of
Fig.~\ref{fig:bifurcationLines}. 
We can see parametric instability suppression as evidenced by the delayed
increase in the amplitude of the spectral peak at $\omega$.
The sharp increase in this peak only occurs well past the parametric instability
transition of Fig.~\ref{fig:bifurcationLines}, when there is no external drive ($F_s=0$).
One sees that the higher $F_s$ is the more suppression in parametric instability
there is.

In Figs. \ref{fig:freqCombPeaks1}-\ref{fig:freqCombPeaks2}, $\omega=1.0625$ and we
vary $F_p$ along line $(b)$ of the bifurcation diagram of Fig.~\ref{fig:bifurcationLines}. 
One can see that the sharp increase in the spectral peak at $\omega$ of the DA
response generates an even stronger frequency-comb-like behavior, with twice as
many peaks, now spaced out in frequency from one another by $\delta$.
By comparing the results of Figs.~\ref{fig:freqCombPeaks1} and
\ref{fig:freqCombPeaks2}, one sees that with increasing value of $F_s$ there is
also a suppression of parametric instability. 
In this case, the parametric instability transition still occurs
in the bistable region, below the threshold to instability in the parametric
oscillator (dashed line).
Apparently, counterintuitively, the larger $F_s$ (the larger the external
excitation) is, the more stable the smaller
amplitude of the parametrically-driven Duffing oscillator response is.

\FloatBarrier

\FloatBarrier
\FloatBarrier

\FloatBarrier
\section{Conclusion}
Here we investigated a one-degree-of-freedom parametrically-driven Duffing
oscillator with a small added ac drive that could present suppression of
parametric instability or present spectral peaks similar to recent experimental
results of mechanical frequency combs \cite{ganesan2017phononic,
ganesan2018phononic, czaplewski2018bifurcation, singh2020giant}.
We have seen two types of frequency comb behavior: one weaker and one stronger.
The stronger frequency comb has twice as many peaks as the weaker form with
peaks at half the distance from one another. 
We claim that the fundamental cause of the frequency-comb dynamical behavior is
a symmetry breaking bifurcation that occurs near the parametric instability
transition.
We have also shown that the averaging method can capture this spectral response.
The stronger form of the frequency comb arises after a subcritical
pitchfork bifurcation of the averaged system of equations of the DA.
In addition, the simple model we propose here could be used as a theoretical
framework, or a toy model, for the study of the frequency comb
phenomenon in mechanical oscillators. 

Furthermore, we point out that our theory is similar to the one developed by K.
Wiesenfeld and McNamara in the 80's for amplification of small signals near
bifurcation points. 
Their theory presented in Refs.~\cite{wies85, wies86} is a linear response
theory based on Floquet theory, whereas here we present a nonlinear response
theory based on the averaging method.
Also, we construct an approximate analytical solution, whereas their model is
generic. One would still have the  difficult task of obtaining the Floquet eigenfunctions.
In addition, the perturbing terms in their prototype example is a parametric
drive, whereas in our models the perturbing terms are the added ac signals.

It is worth mentioning that Bryant and Wiesenfeld
\cite{bryant1986suppression} (see their Fig. 12) obtained an effect similar to the frequency-comb spectral peaks seen here. 
There are several important differences between our physical systems. 
Their Duffing oscillator is not driven parametrically and has a very low quality
factor, which effectively reduces its dimensionality.
They present a suppression of a period-doubling bifurcation, whereas here we
have a suppression of parametric instability due to the small added external ac
signal.
We saw that the suppression increases with the ac signal amplitude.
In our system, the frequency-comb behavior occurs more strongly in the 
region III of the bifurcation diagram, where quiescent solution and limit cycle
are both stable.
Before the jump in amplitude of the central peak, the peaks are spaced by 
$2\delta$, after the jump the peaks are spaced by $\delta$.
The frequency-comb-like spectrum is stronger in a narrow region after the jump of the central
peak.

We investigate one simple nonlinear dynamical system to support our
claim.
The proposed system does not present mode coupling and, thus, has lower dimensionality than the phenomenological models used to explain
the spectral signatures of the experimental mechanical frequency combs.
We believe this could lead to further research to find simpler apparatuses, with
less parameters and less normal modes, that could deliver a frequency-comb-like behavior.
We also believe this work could spur more theoretical work to help understand
better the connection between the frequency comb behavior and the various
bifurcations that occur in conjunction.
For example, in Ganesan \etal \cite{ganesan2018phononic}, one can see
clearly in their Fig. 2 a spectral map that we think has the footprints of a
symmetry-breaking bifurcation as we have seen here. Just below half-way their
spectral map, as the pump amplitude is increased, twice as many peaks appear,
with each new peak in the middle of two pre-existent ones.
Finally, we point out that the equations of the model proposed by Singh \etal
\cite{singh2020giant} for describing their mechanical frequency comb
can be recast in a form similar to Eq.~\eqref{mediaVAm1} if
one integrates out the degrees of freedom of the second normal mode.

%
\end{document}